# Solid friction between soft filaments

Andrew Ward[1], Feodor Hilitski[1], Walter Schwenger[1], David Welch[2], A. W. C. Lau[3], Vincenzo Vitelli[4], L. Mahadevan[5], Zvonimir Dogic[1]

[1] Martin Fisher School of Physics, Brandeis University, 415 South Street, Waltham, MA 02454, USA

[2] Graduate Program in Biophysics and Structural Biology, Brandeis University, 415 South Street, Waltham, MA 02454, USA

[3] Department of Physics, Florida Atlantic University, 777 Glades Road, Boca Raton, FL 33431, USA

[4] Instituut-Lorentz for Theoretical Physics, Universiteit Leiden, 2300 RA Leiden, The Netherlands

[5] School of Engineering and Applied Sciences and Department of Physics, Harvard University, Cambridge, Massachusetts 02138, USA, 29 Oxford Street, Cambridge, MA 02138, USA

**Any macroscopic deformation of a filamentous bundle is necessarily accompanied by local sliding and/or stretching of the constituent filaments. Yet the nature of the sliding friction between two aligned filaments interacting through multiple contacts remains largely unexplored. Here, by directly measuring the sliding forces between two bundled F-actin filaments, we show that these frictional forces are unexpectedly large, scale logarithmically with sliding velocity as in solid-like friction, and exhibit complex dependence on the filaments' overlap length. We also show that a reduction of the frictional force by orders of magnitude, associated with a transition from solid-like friction to Stokes' drag, can be induced by coating F-actin with polymeric brushes. Furthermore, we observe similar transitions in filamentous microtubules and bacterial flagella. Our findings demonstrate how altering a filament's elasticity, structure and interactions can be used to engineer interfilament friction and thus tune the properties of fibrous composite materials.**

Filamentous bundles are a ubiquitous structural motif used for assembly of diverse synthetic, biomimetic and biological materials[1-4]. Any macroscopic deformation of such bundles is necessarily accompanied by local sliding and/or stretching of the constituent filaments[4,5].

Consequently, the frictional forces that arise due to interfilament sliding are an essential determinant of the overall mechanical properties of filamentous bundles. Here, we measure frictional forces between filamentous actin (F-actin) which is an essential building block of diverse biological and biomimetic materials. We bundle F-actin filaments by adding non-adsorbing polymer Poly(Ethylene Glycol) (PEG). As two filaments approach each other, additional free-volume becomes available to PEG coils, leading to the effective attractions interactions, known as the depletion interaction in physics and chemistry and macromolecular crowding in biology ( Supplementary Fig. 1a)[6]. Besides radial interactions, the depletion mechanism also leads to interactions along the filaments' long axis. While the former have been extensively studied using osmotic stress techniques[7], little is known about equally important sliding interactions.

To measure sliding interactions we bundle a pair of actin filaments. Each filament is attached to a gelsolin coated micron-sized bead. Such beads bind exclusively to F-actin barbed end, thus determining the attached filament polarity. Two filaments are held together by attractive depletion forces; subsequently, bead 2 is pulled at a constant velocity with an optical trap while force on bead 1 is simultaneously measured (Fig. 1a, b, Supplementary Movie S1). At first, the force increases as the thermally induced filament slack is pulled out. Subsequently, the force reaches a plateau and thereafter remains constant even as the interfilament overlap length changes by many microns (Fig. 1c). Finally, as the overlap length becomes smaller than a characteristic length scale, the frictional force decreases exponentially and vanishes as the two filaments unbind. Increasing the sliding velocity yields a similar force profile, the only difference being a slightly elevated plateau force $F_{max}$. Repeating these experiments at different velocities reveals that $F_{max}$ exhibits logarithmic dependence on the sliding velocity (Fig. 1d). The strength and range of the attractive depletion potential is tuned by changing polymer concentration and size respectively. This feature allows us to directly relate interfilament sliding friction to cohesive interactions, by simply changing PEG concentration. Stronger cohesion leads to larger plateau force, $F_{max}$ (Fig. 1d).

These experiments reveal several notable features of sliding friction between a pair of F-actin filaments. First, even for the weakest cohesion strength required for assembly of stable bundles, the frictional force is several pN's, comparable to the force exerted by myosin motors. Second, above a critical value, the frictional force is independent of the interfilament overlap length. Third, the plateau force, $F_{max}$, exhibits a logarithmic dependence on the sliding velocity. These observations are in sharp contrast with models that approximate biopolymers as a structureless filament interacting through excluded volume interactions. Frictional coupling between such homogeneous filaments would be dominated by hydrodynamic interactions, resulting in forces that are linearly dependent on both the pulling velocity and overlap length, and orders of magnitude weaker than those measured. Since these features are not observed experimentally we exclude hydrodynamic interactions as a dominant source of frictional coupling and reconsider the basic physical processes at work.

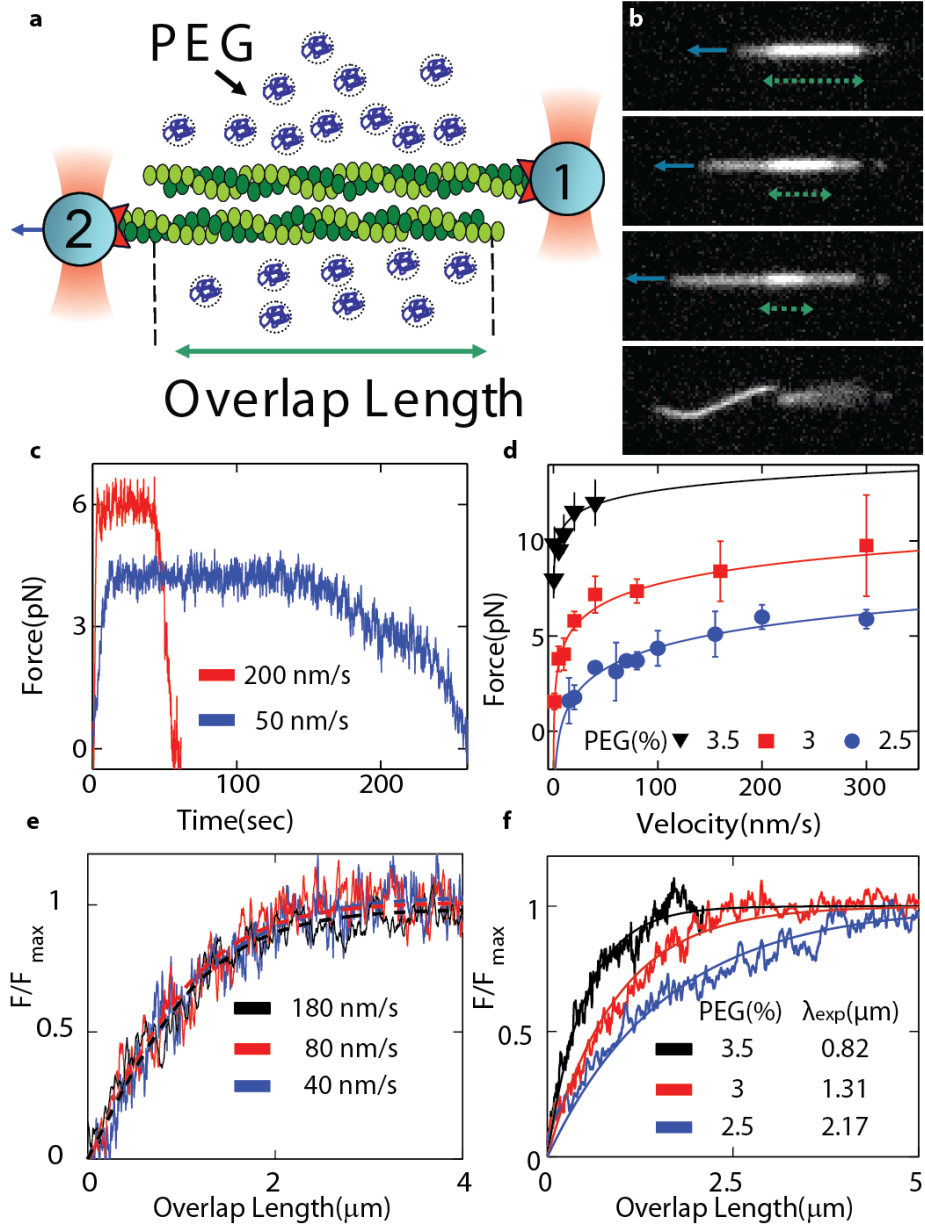

**Figure 1.** Single-molecule experiments reveal frictional interactions between a pair of sliding F-actin filaments. **(a)** Schematic of the experimental setup. Actin filaments attached to gelsolin coated beads are assembled into anti-parallel bundles using optical traps. Bead 2 is pulled at a constant velocity while simultaneously measuring the force exerted on bead 1. **(b)** A sequence of images illustrates two filaments being pulled apart. The green dashed line indicates the interfilament overlap length and the blue arrow indicates the pulling direction (Supplementary Movie S1). **(c)** Time dependence of the frictional force measured for two pulling velocities. **(d)** The frictional force, $F_{max}$, exhibits logarithmic dependence on the pulling velocity. The measurements are repeated at three different cohesion strengths (PEG concentrations). Lines indicate fits of Eq. 2 to experimental data. **(e)** The dependence of the frictional force $F/F_{max}$ on the interfilament overlap length. Force profiles taken at different sliding velocities rescale onto a universal curve (Eq. 1), defining a velocity independent frictional kink width $\lambda$. **(f)** Stronger cohesion leads to smaller $\lambda$. For all experiments, the salt concentration is 200 mM KCl.

Certain aspects of the frictional interactions between actin filaments can be understood by studying the sliding dynamics of two commensurate 1D lattices of beads and springs under shear (Fig. 2a). The lattices do not slide past each other rigidly. Instead, the mechanism of sliding involves the propagation of localized excitations called kinks that carry local compression of the lattice (Fig. 2b). Every time a kink propagates across the filament the two intercalating lattices slide by one lattice spacing. Sliding happens locally, yielding a frictional force that is controlled by the kink width, $\lambda$, rather than the total overlap length, $L$, provided that $L \gg \lambda$, as is typically the case in conventional friction. However, in our experiments the filament overlap can be controlled from nanometers to many microns allowing us to examine the regime where $L \leq \lambda$. In this regime, a propagating kink cannot fully develop and the sliding force exhibits a dependence on $L$.

These arguments can be quantitatively rationalized within the framework of the Frenkel-Kontorova model[8,9]. The sliding filament is modeled as a 1D lattice of length $L$ comprising beads connected by springs with stiffness constant, $k$. The lattice periodicity is given by the actin monomer spacing, $d$. The interaction with the stationary filament is modeled by a commensurate sinusoidal background potential of depth, $U_0$, and periodicity $d$. In the continuum limit the bead displacement field, $u(x)$ satisfies the Sine-Gordon equation, $\lambda^2 u_{xx} = \sin\left(\frac{u(x)}{d}\right)$, which admits a static kink solution of the form $u_s(x) = 4d \tan^{-1}\left(e^{-\frac{x}{\lambda}}\right)$, where $\lambda$ is the kink width that corresponds to the length of the lattice that is distorted (see Supplementary Methods). Kink width is determined by the ratio of filament stiffness to the stiffness of the background potential: $\lambda^2 = kd^4/U_0$. For very stiff filaments such as F-actin an imposed distortion will extend over many lattice spacings.

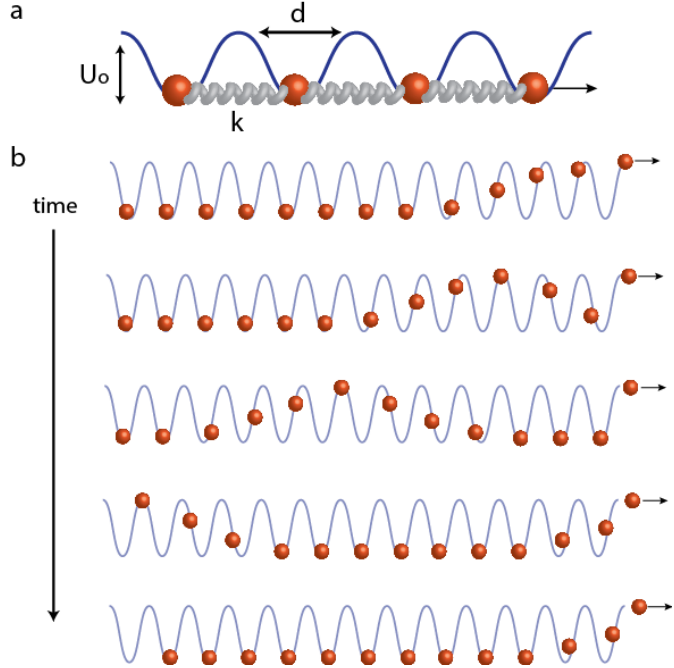

**Figure 2.** 1D-Frenkel-Kontorova model accounts for the essential features of interfilament sliding friction. **(a)** Schematic of a model in which a sliding filament is approximated as a periodic lattice of points connected by stiff springs. A sinusoidal background potential models the interaction with the stationary filament. **(b)** Schematic of how a filament slides by one lattice spacing in a response to an applied pulling force (Supplementary Movie S2). For filaments with finite extensibility the applied force decays over a characteristic lengthscale that is determined by the ratio of spring stiffness and the stiffness of the background potential. The first bead hoping over the rightmost barrier is accompanied by soliton formation that propagates leftwards. Once the soliton reaches the leftmost bead the entire filament is translated by a lattice spacing $d$.

We first consider the case $L \leq \lambda$ and assume that the finite size chain located from x=-L to x=0 is gradually pulled out at x=0. The pulling force, $F$, displaces the right-most bead to the maximum of the potential (i.e. $u(0)=d/2$) generating a strain field, $u_x(\mathrm{x})$, that decays exponentially inside the sample with a characteristic length $\lambda$. Once the right-most bead hops over the maximum in the potential, the whole lattice slides by $d$ and every bead falls to the bottom of the respective potential generating a state of vanishing strain and energy. Repeating this process $n$ times translates the left-most edge from $-L$ to $-L+nd$. The work done by the pulling force in the $n^{th}$ run, $F(n)\,d$, is the energy difference between the elastic energy stored in the kink configuration $u_s(x)$ and the uniform state $u(x)=0$, which has zero energy. Since the elastic energy is proportional to $F_{max} \, d \, \text{Sech}^2(x/\lambda)$ over the part of the chain still interacting with the background potential (i.e. from $x=L+nd$ to $x=0$), the resulting force reads:

$$F(n) = F_{max} \tanh\left(\frac{L-nd}{\lambda}\right) \qquad (\text{Eq. 1}),$$

where $L$-$nd$ represents the remaining overlap length between the two filaments. If the overlap length is larger than the kink width ($L > \lambda$), $F(n)$ saturates at $F_{max}$. In this limit a kink nucleated at the right-most edge can fully develop and propagate down the chain, progressively shifting the particles it leaves behind (see Supplementary Fig. 2. and Movie S2). As a result the force ceases to depend on the overlap length. Instead it is set by the kink width, which remains equal to the static value $\lambda$, unaffected by the kink dynamics in the overdamped regime (see Supplementary Information).

Experiments reveal that $F_{max}$ scales logarithmically with the pulling speed, $v$ (Fig. 1d). The intuitive reason for this dependence is that as the lattice is pulled, the particles within the kink undergo thermally assisted hopping through the periodic background potential. A classical model, originally formulated by Prandtl and Thomlinson, predicts:

$$F_{max} = \frac{k_B T}{d} \log\left(\frac{v}{2df_c} e^{U_0/k_B T}\right) \qquad (\text{Eq.2})$$

where $T$ is the temperature and $1/f_c$ is the relaxation time of a monomer in a potential energy well[10-13]. Fitting Eq. 2 to the plateau value of the force-velocity curves reveals that the periodicity of the background potential is ~5 nm (Supplementary Table S1), in quantitative agreement with F-actin monomer spacing[14]. This result suggests that cohering F-actin monomers intercalate with each other, and sliding interactions require monomers to either deform or hop over each other.

Eq. 1 predicts that the force profiles taken at varying pulling speeds should fall onto a master curve once rescaled by $F_{max}(v)$. This data collapse is demonstrated in Fig. 1e (Supplementary Table S2). It yields experimental measurement of velocity-independent kink width, $\lambda$, which we compare to the theoretical prediction $\lambda = \sqrt{\frac{kd^4}{U_0}}$. We take the lattice periodicity to be 5.5 nm[14] and $k$ to be ~7000 pN/nm[15]. To estimate $U_o$ we measure the strength of the depletion induced attraction by allowing an isolated filament to fold into a racquet-like configuration (Supplementary Fig. 1c, d)[16]. The size of the racquet head is directly related to the filament cohesion strength per unit length, $U_o$. Without any adjustable parameters our theoretical model predicts values of $\lambda$ which are of the same order of magnitude as those extracted from experiments (Supplementary Table S3). Increasing depletant concentration increases $U_o$, leads to a decrease in $\lambda$, which is again in agreement with the theoretical prediction. In summary, we have demonstrated that the tunable kink width critically determines the dependence of frictional force on the overlap length between the two intercalating nanofilaments. A lengthscale similar to $\lambda$ arises in many other materials science contexts, such as shearing of double stranded DNA[17,18].

Previous experiments have uncovered directionally dependent friction in both biological and synthetic materials[19-21]. To investigate the directional dependence of interfilament sliding friction between polar actin filaments we have altered the experimental configuration by attaching beads to both ends of one filament (Supplementary Fig. 3). Using this configuration we find that $F_{max}$ for sliding anti-parallel filaments is approximately twice as large compared to filaments with parallel alignment (Fig. 3a, b). While $F_{max}$ is different, the scaling of frictional force with velocity and filament overlap length is the same for both orientations indicating that same physics describes sliding of both polar and anti-polar filaments. Furthermore, we also investigate stress relaxation upon an application of a step strain (Fig. 3c). For parallel configuration the applied stress quickly relaxes to a finite but small force. In contrast, for anti-parallel orientation the applied stress relaxes on much longer time scales. These experiments indicate that the axial interaction potential between sliding F-actin filaments is polar and thus sliding actin filaments can act as molecular ratchets (Fig. 3d).

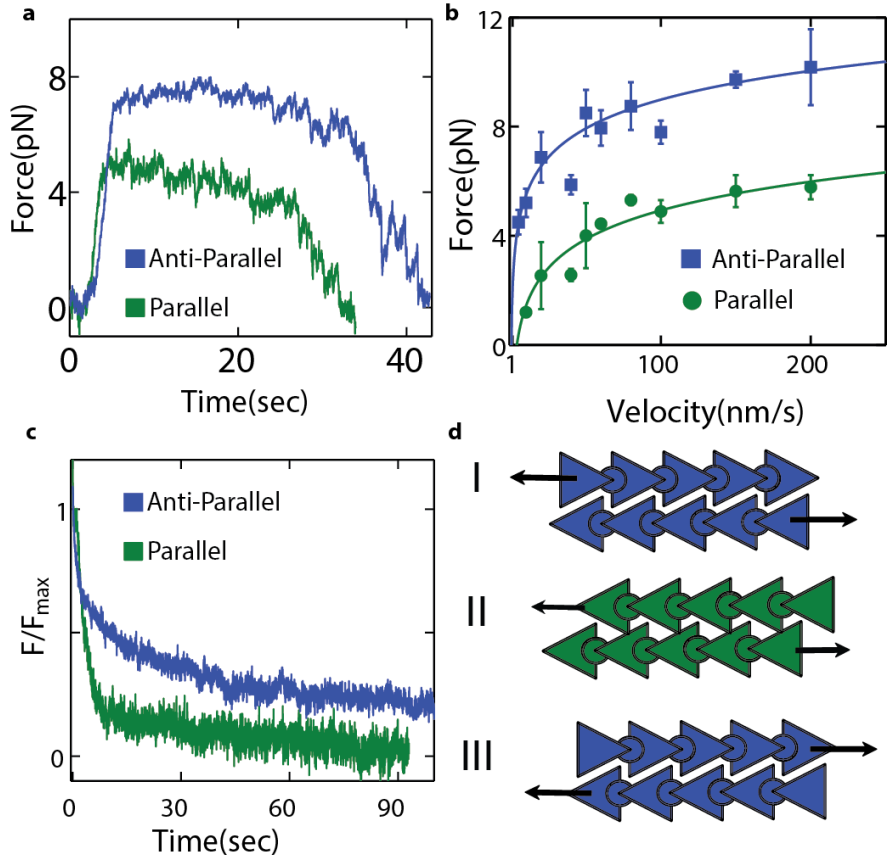

**Figure 3.** Interfilament sliding friction depends on relative filament polarity. **(a)** Two profiles describing time dependent frictional force. The only difference is the relative filament polarity. **(b)** The plateau force, $F_{max}$, exhibits pronounced directional dependence. **(c)** Relaxation of a force upon application of a step strain. For parallel filaments the force quickly relaxes to a small but finite value. For anti-parallel filaments the force relaxes on much longer timescales. **(d)** Schematic representation of actin filaments that account for the directional dependence of sliding friction. Anti-parallel and parallel arrangements correspond to schemes I and II, respectively. We are not able to measure friction for configuration III. All experiments were performed at salt concentration of 400 mM KCl.

Armed with a basic understanding of filament sliding friction we next devise practical methods to tune its magnitude. One possible method to accomplish this is by changing filament structure. We decorated F-actin with a covalently attached PEG brush. In this system, friction was quantified by visualizing sliding dynamics of bundled filaments. Native F-actin bundles did not exhibit any thermally driven sliding in agreement with our previous measurements (Fig. 4a). In contrast, PEG coated F-actin formed bundles in which individual filaments freely slid past each other due to thermal fluctuations (Fig. 4b, Supplementary Movie S3). To extract quantitative data, we measure the mean square displacement (MSD) of the relative position of the short filament with respect to the longer filament to which it is bound (Fig. 4c). The linear MSD curves are consistent with hydrodynamic coupling between PEG coated filaments; the slope yields the diffusion of a bound filament which is 5-fold smaller than that of an isolated filament[22]. It follows, that the hydrodynamic friction coefficient of a bundled 5 μm long filament is ~ $10^{-4}$ pN s/nm. Pulling such a filament at 100 nm/sec would result in a 10 fN force, which is three orders of magnitude smaller than the comparable forces measured for bare F-actin. This demonstrates how simple structural modifications greatly alter the sliding filament friction. We compare these results with those for another important biopolymer, a microtubule. Previous study has shown that unlike F-actin sliding friction of microtubule is weak and dominated by hydrodynamic interactions (Fig. 4d)[23]. Microtubule surfaces are coated with a charged disordered amino-acid domain known as e-hooks[24]. We hypothesize that these domains might act as an effective polyelectrolye brush, screening molecular interactions and thus lowering sliding friction. To test this hypothesis we remove e-hooks with an appropriate protease. When treated in such a way microtubule bundles do not exhibit any interfilament sliding indicating much higher sliding friction (Fig. 4e,f). Such observations agree with previous studies which have shown that brush-like surfaces can drastically lower friction coefficients[25].

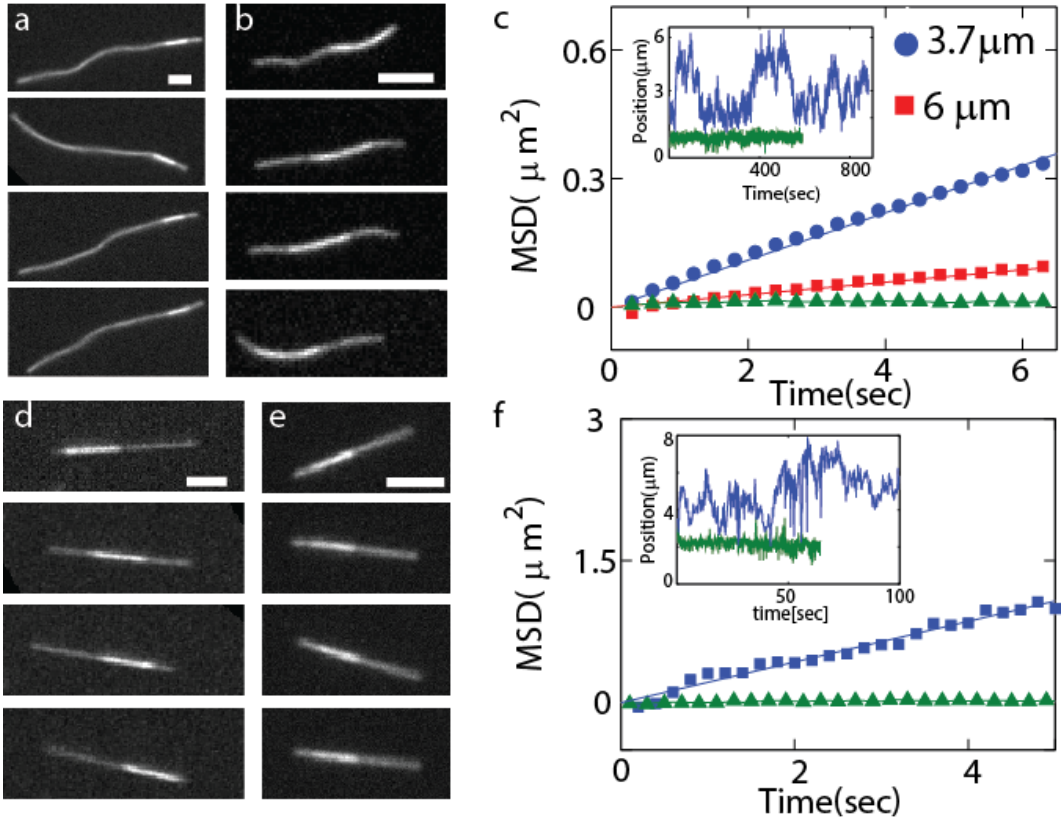

**Figure 4.** Filament surface structure controls the transition from solid to hydrodynamic friction. **(a)** A sequence of images illustrates the relative diffusion of a bundled pair of unmodified F-actin filaments. The brighter region, indicating the filament overlap area, is frozen at a specific location due to absence of any thermally induced filament sliding. **(b)** F-actin filaments coated with a PEG polymer brush exhibit thermally driven sliding, due to a significantly reduced frictional coupling (Supplementary Movie S3). **(c)** For PEG-coated bundles, mean square displacement (MSD) of the short filament with respect to the longer filament increases linearly with time, indicating a hydrodynamic coupling. Inset: Relative position of a filament diffusing within the bundle for both coated (blue) and uncoated (green) F-actin. **(d)** Untreated microtubule bundles exhibit diffusive sliding that is dominated by hydrodynamic coupling. **(e)** Removing brush like e-hooks from microtubule surface leads to bundles that exhibit no sliding. **(f)** MSDs of a bundle of microtubules (blue) compared with a bundle of subtilisin treated microtubules (green). Inset: Relative position of a microtubule bundle for both non-subtilisin (blue) and subtilisin treated microtubule (green). Scale bars, 3 μm.

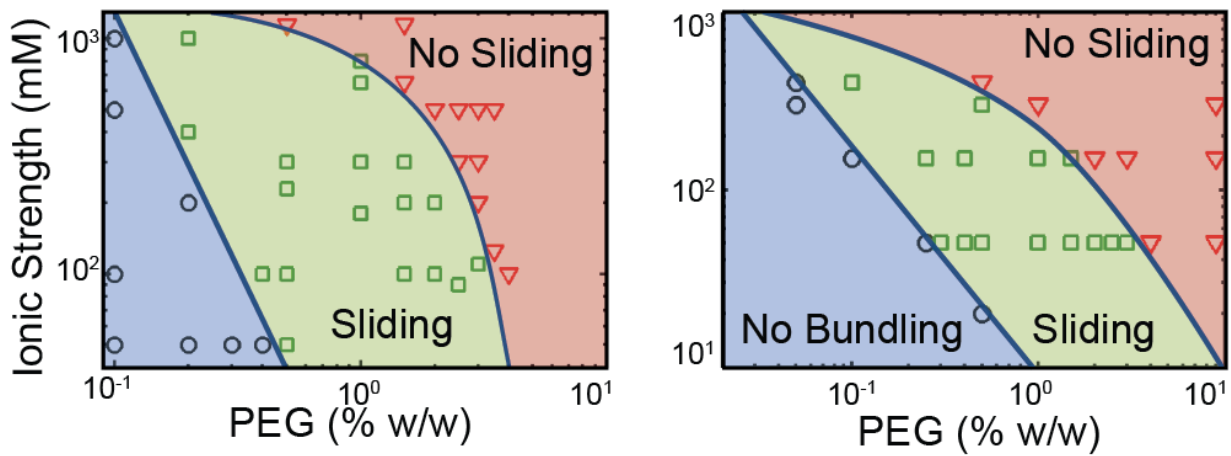

**Figure 5.** Sliding dynamics of MTs and bacterial flagella. **(a)** Microtubule sliding dynamics depends on the depletant concentration and suspension ionic strength. At low depletant concentration and ionic strength filaments remain unbundled. To induce bundle formation it is necessary to either increase depletant concentration or decrees electrostatic screening length. In this regime bundles exhibit sliding dynamics. With increasing ionic strength or depletant concentration sliding filaments undergo a sharp crossover into a state with no detectable sliding indicating a stronger frictional coupling. **(b)** Straight bundled flagellar filaments exhibit sliding similar to that of microtubule bundles.

Alternatively frictional coupling can be tuned by engineering lateral interfilament interactions. We examined how sliding dynamics of three different filaments (F-actin, microtubules, bacterial flagella) depends on the strength of lateral filament attraction, which is controlled by the depletant concentration, and average filament separation, which is tuned by the ionic strength (Fig. 5). Microtubules and bacterial flagella exhibited two distinct dynamical states. For low depletant concentration (weak attraction) and low ionic strength the filaments have large lateral separation and freely slide past each other. Such dynamics indicates weak frictional coupling that is dominated by hydrodynamic interactions. Increasing depletant concentration or ionic strength above a critical threshold induces a sharp transition into a distinct dynamical state that exhibits no measurable sliding even after tens of minutes of observation time. The sliding dynamics of flagella and microtubules are remarkably similar to each other. In comparison, native F-actin filaments only displayed a non-sliding state indicative of solid-like friction, for all parameters explored. The observation of non-sliding dynamics in three structurally diverse filaments suggests that solid-like frictional coupling is a common feature of biological filaments.

To summarize, the mechanical properties of composite filamentous bundles are not only determined by the rigidity of the constituent filaments but also by their inter-filament interactions, such as the cohesion strength and sliding friction[26]. Therefore, quantitative models of composite bundle mechanics must account for inter-filament sliding friction. We have demonstrated an experimental technique that enables measurement of such forces. We directly measured frictional forces between chemically identical F-actin filaments thus bridging the gap between previous studied friction of sliding point-like contacts[27,28] and 2D surface[29-31]. Combining such measurements with simulations and theoretical modeling we have described design principles required to engineer inter-filament friction and thus tune properties of frictionally interacting composite filamentous materials.

**Methods:** Actin and Gelsolin were purified according to previously published protocols. Actin filaments were polymerized in high salt buffer and subsequently stabilized with Alexa-488-phalloidin. Gelsolin was covalently coupled to one micron carboxylic coated silica beads by 1-Ethyl-3-[3-dimethylaminopropyl]carbodiimide hydrochloride. All experiments were performed in buffer suspension containing 20mM Phosphate at pH=7.5, 300mg/mL sucrose and either 200mM or 400mM KCl. Poly (Ethylene Glycol) (MW 20,000 Da) was used as a depletion agent at concentrations ranging between 20-35 mg/mL as indicated in the manuscript. An optical trap (1064 nm) was time shared between multiple positions using Acousto-optic Deflector. One bead was translated at a constant velocity while the force exerted on the other bead was measured using back focal plane interferometry technique. Simultaneously, images of sliding F-actin filaments were acquired using fluorescence microscopy on a Nikon Eclipse Te2000-u microscope equipped with Andor iKon-M CCD. Sliding dynamics was visualized by confining a two-filament bundle into a quasi-2D microscope chamber thus ensuring that filaments stay in focus. The surfaces were coated with poly-acrylamide brush, which suppresses adsorption of filaments. Bacterial flagella were isolated from *Salmonella typhimurium* strain SJW 605. The flagellin protein of these bacteria has a mutation that causes flagellar filaments to assume a straight shape. Microtubules were isolated following standard protocols.

**Acknowledgments:** We acknowledge useful discussions with Nitin Upadhyaya, M. Hagan, and R. Bruinsma. AW, DW and WS were supported by National Science Foundation grants CMMI-1068566, NSF-MRI-0923057 and NSF-MRSEC-1206146. FH and ZD were supported by Department of Energy, Office of Basic Energy Sciences under Award DE-SC0010432TDD. VV acknowledges FOM and NWO for financial support. LM was supported by Harvard-NSF

MRSEC and MacArthur Foundation. We also acknowledge use of Brandeis MRSEC optical microscopy facility (NSF-MRSEC-1206146).

**Author contributions:** A.W. and Z.D. conceived the experiments. A.W. measured actin sliding friction and performed computer simulations. F.H, A.W. and D.W. performed microtubule sliding experiments. W.S. performed flagella sliding dynamics. A. W. C. L. developed preliminary theoretical model that explains velocity dependence of sliding friction. L.M. and V.V. developed the theoretical model that explains dependence of sliding friction on overlap length. A.W., V.V. and Z. D. wrote the manuscript. All authors revised the manuscript.

**Additional information:** Supplementary information is available in the online version of the paper. Reprints and permissions information is available online at www.nature.com/reprints. Correspondence and requests for materials should be addressed to Z.D.

**Competing financial interests:** The authors declare no competing financial interests.

**References:**

1. Vigolo, B. *et al.* Macroscopic fibers and ribbons of oriented carbon nanotubes. *Science* **290**, 1331-1334, doi:10.1126/science.290.5495.1331 (2000).
2. Sanchez, T., Chen, D. T. N., DeCamp, S. J., Heymann, M. & Dogic, Z. Spontaneous motion in hierarchically assembled active matter. *Nature* **491**, 431-+, doi:10.1038/nature11591 (2012).
3. Kozlov, A. S., Baumgart, J., Risler, T., Versteegh, C. P. C. & Hudspeth, A. J. Forces between clustered stereocilia minimize friction in the ear on a subnanometre scale. *Nature* **474**, 376-379, doi:10.1038/nature10073 (2011).
4. Claessens, M., Bathe, M., Frey, E. & Bausch, A. R. Actin-binding proteins sensitively mediate F-actin bundle stiffness. *Nature Materials* **5**, 748-753, doi:10.1038/nmat1718 (2006).
5. Heussinger, C., Bathe, M. & Frey, E. Statistical mechanics of semiflexible bundles of wormlike polymer chains. *Physical Review Letters* **99**, doi:10.1103/PhysRevLett.99.048101 (2007).
6. Zimmerman, S. B. & Minton, A. P. Macromolecular crowding - biochemical, biophysical and physiological consequences *Annual Review of Biophysics and Biomolecular Structure* **22**, 27-65, doi:10.1146/annurev.bb.22.060193.000331 (1993).
7. Rau, D. C., Lee, B. & Parsegian, V. A. Measurement of the repulsive force between poly-electrolyte molecules in ionic solution - hydration forces between parallel DNA double helices. . *Proceedings of the National Academy of Sciences of the United States of America-Biological Sciences* **81**, 2621-2625, doi:10.1073/pnas.81.9.2621 (1984).
8. Vanossi, A., Manini, N., Urbakh, M., Zapperi, S. & Tosatti, E. Colloquium: Modeling friction: From nanoscale to mesoscale. *Rev. Mod. Phys.* **85**, 529-552, doi:10.1103/RevModPhys.85.529 (2013).
9. Braun, O. M. & Kivshar, Y. S. Nonlinear dynamics of the Frenkel-Kontorova model. *Physics Reports-Review Section of Physics Letters* **306**, 1-108, doi:10.1016/s0370-1573(98)00029-5 (1998).
10. Merkel, R., Nassoy, P., Leung, A., Ritchie, K. & Evans, E. Energy landscapes of receptor-ligand bonds explored with dynamic force spectroscopy. *Nature* **397**, 50-53 (1999).
11. Gnecco, E. *et al.* Velocity dependence of atomic friction. *Physical Review Letters* **84**, 1172-1175, doi:10.1103/PhysRevLett.84.1172 (2000).
12. Muser, M. H., Urbakh, M. & Robbins, M. O. in *Advances in Chemical Physics, Vol 126* Vol. 126 *Advances in Chemical Physics* (eds I. Prigogine & S. A. Rice)  187-272 (2003).
13. Suda, H. Origin of friction derived from rupture dynamics. *Langmuir* **17**, 6045-6047, doi:10.1021/la0106384 (2001).
14. Holmes, K. C., Popp, D., Gebhard, W. & Kabsch, W. Atomic model of the actin filament *Nature* **347**, 44-49, doi:10.1038/347044a0 (1990).
15. Kojima, H., Ishijima, A. & Yanagida, T. Direct measurementof stifness of single actin-filaments with and wihtout topomyosin by in-vitro nanomanipulation. *Proceedings of the National Academy of Sciences of the United States of America* **91**, 12962-12966, doi:10.1073/pnas.91.26.12962 (1994).
16. Lau, A. W. C., Prasad, A. & Dogic, Z. Condensation of isolated semi-flexible filaments driven by depletion interactions. *Epl* **87**, doi:4800610.1209/0295-5075/87/48006 (2009).

17    De Gennes, P. G. Maximum pull out force on DNA hybrids. *Comptes Rendus De L Academie Des Sciences Serie Iv Physique Astrophysique* **2**, 1505-1508, doi:10.1016/s1296-2147(01)01287-2 (2001).
18    Hatch, K., Danilowicz, C., Coljee, V. & Prentiss, M. Demonstration that the shear force required to separate short double-stranded DNA does not increase significantly with sequence length for sequences longer than 25 base pairs. *Physical Review E* **78**, doi:01192010.1103/PhysRevE.78.011920 (2008).
19    Bormuth, V., Varga, V., Howard, J. & Schaffer, E. Protein Friction Limits Diffusive and Directed Movements of Kinesin Motors on Microtubules. *Science* **325**, 870-873, doi:10.1126/science.1174923 (2009).
20    Choi, J. S. *et al.* Friction Anisotropy-Driven Domain Imaging on Exfoliated Monolayer Graphene. *Science* **333**, 607-610, doi:10.1126/science.1207110 (2011).
21    Forth, S., Hsia, K.-C., Shimamoto, Y. & Kapoor, T. M. Asymmetric Friction of Nonmotor MAPs Can Lead to Their Directional Motion in Active Microtubule Networks. *Cell* **157**, doi:10.1016/j.cell.2014.02.018 (2014).
22    Li, G. L. & Tang, J. X. Diffusion of actin filaments within a thin layer between two walls. *Physical Review E* **69**, doi:06192110.1103/PhysRevE.69.061921 (2004).
23    Sanchez, T., Welch, D., Nicastro, D. & Dogic, Z. Cilia-Like Beating of Active Microtubule Bundles. *Science* **333**, 456-459, doi:10.1126/science.1203963 (2011).
24    Nogales, E., Wolf, S. G. & Downing, K. H. Structure of the alpha beta tubulin dimer by electron crystallography. *Nature* **391**, 199-203, doi:10.1038/34465 (1998).
25    Klein, J., Kumacheva, E., Mahalu, D., Perahia, D. & Fetters, L. J. Reduction of frictional forces between solid-surfaces bearing polymer brushes. *Nature* **370**, 634-636, doi:10.1038/370634a0 (1994).
26    Akbulut, M., Belman, N., Golan, Y. & Israelachvili, J. Frictional properties of confined nanorods. *Advanced Materials* **18**, 2589-+, doi:10.1002/adma.200600794 (2006).
27    Mate, C. M., McClelland, G. M., Erlandsson, R. & Chiang, S. Atomic scale-friction of a tungsten tip on a graphite surface. *Physical Review Letters* **59**, 1942-1945, doi:10.1103/PhysRevLett.59.1942 (1987).
28    Luan, B. Q. & Robbins, M. O. The breakdown of continuum models for mechanical contacts. *Nature* **435**, 929-932, doi:10.1038/nature03700 (2005).
29    Yoshizawa, H., Chen, Y. L. & Israelachvili, J. Fundamental mechanism of interfacial friction. 1. Relation between adhesion and friction. *Journal of Physical Chemistry* **97**, 4128-4140, doi:10.1021/j100118a033 (1993).
30    Van Alsten, J. & Granick, S. Molecular tribometry of ultrathin liquid-films. *Physical Review Letters* **61**, 2570-2573, doi:10.1103/PhysRevLett.61.2570 (1988).
31    Bohlein, T., Mikhael, J. & Bechinger, C. Observation of kinks and antikinks in colloidal monolayers driven across ordered surfaces. *Nature Materials* **11**, 126-130, doi:10.1038/nmat3204 (2012).